\documentclass[aps,twocolumn,showpacs]{revtex4}
\usepackage[dvips]{graphics}

\begin{document}

\title{Ladder network as a mesoscopic switch: An exact result}

\author{Shreekantha Sil$^1$, Santanu K. Maiti$^2$ and Arunava Chakrabarti$^3$}

\affiliation{$^1$Department of Physics, Visva-Bharati, Santiniketan, 
West Bengal-731 235, India. \\
$^2$Department of Physics, Narasinha Dutt College, 129 Belilious Road, 
Howrah-711 101, India. \\
$^3$Department of Physics, University of Kalyani, Kalyani, 
West Bengal-741 235, India.} 

\begin{abstract}
We investigate the possibilities of a tight binding ladder network as a
mesoscopic switching device. Several cases have been discussed in
which any one or both the arms of the ladder can assume random, ordered or
quasiperiodic distribution of atomic potentials. We show that, for a special
choice of the Hamiltonian parameters it is possible to prove exactly the
existence of mobility edges in such a system, which plays a central role
in the switching action. We also present numerical results for the 
two-terminal conductance of a general model of a quasiperiodically grown 
ladder which support the general features of the electron states in such 
a network. The analysis might be helpful in fabricating mesoscopic or DNA 
switching devices.
\end{abstract}


\maketitle 

\noindent
Understanding the character of single particle states in low dimensional
quantum systems has always been an interesting problem in condensed matter
theory. It is well known that in one dimension, irrespective of the strength
of disorder, all the single particles states are exponentially
localized~\cite{ander58,lee85}. Later, scaling arguments \cite{tvr} led to
the result that all states should be exponentially localized even in two
dimensions for arbitrarily weak disorder. Mobility edges separating the
extended (conducting) states from the localized (insulating) ones do not
exist in one or two dimensional systems with random disorder. Some exceptions
to this `rule' have of course been suggested in the past in a variation of
the quasiperiodic Aubry-Andre model~\cite{aubry79,eco82,das88,das90,rolf90,
rolf91}, and later, in the so called {\em correlated disordered} models in 
one dimension~\cite{dun90,san94,fabf98,domin03}. However, an analytical
proof of a metal-insulator transition (MIT) is yet to be achieved in
low-dimensions.

\begin{figure}[ht]
{\centering\resizebox*{7cm}{3cm}{\includegraphics{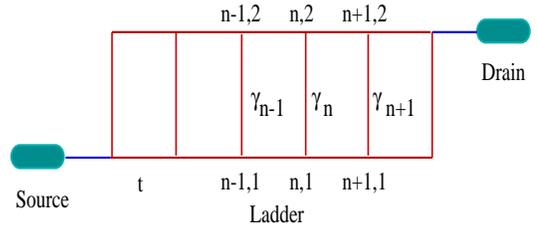}} \par}
\caption{(Color online). Schematic view of a ladder attached to two 
electrodes.}
\label{bridge}
\end{figure}

In this article we investigate the electronic spectrum of a two-chain
ladder network within a tight binding approximation for non-interacting
electrons. This ladder network is built by coupling two one dimensional
chains laterally (see Fig.~\ref{bridge}). The chains may or may not be
identical and, are coupled to each other at every vertex through an
interchain hopping integral. The motivation behind the present work is
twofold. First, we wish to investigate if the quasi-one dimensional
structure of the network, for a suitable combination of the site potentials
and the inter-site hopping integrals, leads to a possibility of observing
an MIT. If it is true, then a ladder network such as
this, could be used as a switching device, the design of which is of great
concern in the current era of nanofabrication. Secondly, the ladder
networks have recently become extremely important in the context of
understanding the charge transport in double stranded DNA~\cite{macia06,
rudo07}. The possibility of observing a localization-delocalization
transition in a DNA-like double chain has already been numerically addressed 
within
a tight binding framework by Caetano and Schulz \cite{cat}. In view of this,
the examination of the electronic spectrum of a ladder network might throw
new light, both in the context of basic physics and possible technological
applications. It may be mentioned that, in a recent article \cite{san} the 
present authors proved the existence of an MIT in an aperiodic Aubry ladder
network. However, the results in that work strongly depend on the dual 
symmetry exhibited by an Aubry model \cite{aubry79}. So, to our mind, 
whether an MIT really exists for a general disordered ladder still remains 
a challenging problem.

We adopt a tight binding formalism and incorporate only the nearest neighbor
hopping inside a plaquette of the ladder. Interestingly, even for a
disordered ladder a certain correlation between the system-parameters
allows us to perform an exact analysis of the energy spectrum and make
definite comments on the character of the single particle states. The
variation in the conductance of the network, which may even exhibit a
crossover from a completely opaque to a fully or partly transmitting one,
is easily understood. In view of such a crossover one can then set the
Fermi energy at a suitable energy zone in the spectrum and control the
transmission characteristics. This enhances the prospect of such ladder
networks as novel switching devices. The possibility of designing DNA
devices, to our mind, can also be encouraged by such analysis. Our results
are exact. Finally, we present numerical results for a quasiperiodically
ordered ladder network by evaluating the two terminal conductance within
a Green's function formalism. The conductance spectrum not only corroborates
the general features of disordered networks discussed previously and as
revealed in our analytical approach, but also shows the presence of
localization-de-localization transition in the quasiperiodic ladder network.
Again in this case, analytical results may be obtained by appropriately
adjusting the system parameters.

Let us refer to Fig.~\ref{bridge}. The Hamiltonian of the ladder
network is given by,
\begin{equation}
{\mathbf H}=\sum_n {\bf {\epsilon_n c_n^{\dagger} c_n}} +
{\mathbf t} \sum_n {\mathbf c_n^{\dagger} c_{n+1}} + h.c.
\label{equ1}
\end{equation}
where,
\begin{equation}
{\bf c_n} = \left(\begin{array}{c} c_{n,1} \\ c_{n,2}\end{array}\right)
\label{equ2}
\end{equation}
In the above, $c_{n,j}$ ($c_{n,j}^{\dagger}$) are the annihilation (creation)
operator at the $n$th site of the $j$th ladder.
\begin{eqnarray}
\bf{\epsilon_n} & = & \left( \begin{array}{cc} \epsilon_{n,1} & \gamma_n \\
\gamma_n & \epsilon_{n,2} \end{array} \right) \nonumber \\
\bf{t} & = & \left( \begin{array}{cc} t & 0 \\
0 & t \end{array} \right)
\label{equ3}
\end{eqnarray}
where $\epsilon_{n,j}$ is the on-site potential at the $n$th site of the
$j$th ladder, $\gamma_n$ is the vertical hopping between the $n$th sites
of the two arms of the ladder and, $t$ is the nearest-neighbor hopping
integral between the $n$th and the ($n+1$)th sites of every arm.

We describe the system in a basis defined by the vector
\begin{equation}
{\bf f_n} = \left(\begin{array}{c} f_{n,1} \\ f_{n,2}\end{array}\right)
\label{equ4}
\end{equation}
where, $f_{n,j}$ is the amplitude of the wave function at the $n$th site of
the $j$th arm of the ladder, $j$ being equal to $1$ or $2$. Using this basis,
our task is to obtain solutions of the difference equation
\begin{equation}
(E{\bf I}-{\bf \epsilon_n})  {\mathbf f_n} = {\bf t}( {\bf f_{n+1}} +
{\bf f_{n-1}} )
\label{equ5}
\end{equation}
{\bf $I$} being the $2 \times 2$ identity matrix. Let us now separately
discuss cases which will throw light on the central problem addressed in
this paper, viz, the possibility of getting a localization-delocalization
transition in such a system.
\vskip 0.05in
\noindent
{\em Case I:} $\epsilon_{n,2} = \epsilon \epsilon_{n,1}$
and, $\gamma_n = \gamma \epsilon_{n,1}$.
\vskip 0.05in
We introduce the above correlation
between the on-site potentials at each arm. The selection of $\epsilon_{n,1}$
is of course, done in a random manner. With this choice of the parameters,
the difference equation ($5$) reads,
\begin{equation}
\left[E{\bf I} - \epsilon_{n,1}{\bf M}  \right] {\bf f_n} = t {\bf I}
\left({\bf f_{n+1}} + {\bf f_{n-1}} \right)
\label{equ6}
\end{equation}
where,
\begin{eqnarray}
\bf{M} & = & \left( \begin{array}{cc} 1 & \gamma \\
\gamma & \epsilon \end{array} \right)
\label{equ7}
\end{eqnarray}

We now diagonalize the matrix $\bf{M}$ by a similarity transformation using
a matrix $\bf{S}$, and define
\begin{equation}
{\bf \phi_n} = {S^{-1}} {\bf f_n}
\label{equ8}
\end{equation}
The above difference equation ($6$) now decouples, in this new basis, in the
following pair of equations,
\begin{equation}
(E-\lambda_1 \epsilon_{n,1}) \phi_{m,1} = t (\phi_{n+1,1} + \phi_{n-1,1})
\label{equ9}
\end{equation}
\begin{equation}
(E-\lambda_2 \epsilon_{n,1}) \phi_{n,2} = t (\phi_{n+1,2} + \phi_{n-1,2})
\label{equ10}
\end{equation}

Here, $\phi_{n,1}$ and $\phi_{n,2}$ are the elements of the column vector
${\bf \phi_n}$ and, $\lambda_1$ and $\lambda_2$ are the eigenvalues of the
matrix $\bf{M}$ which are given by,
\begin{eqnarray}
\lambda_1 & = & \frac{1+\epsilon}{2} + \sqrt {\left[\left (\frac{1-\epsilon}
{2} \right)^2 + \gamma^2 \right]} \nonumber \\
\lambda_2 & = & \frac{1+\epsilon}{2} -
\sqrt {\left[ \left (\frac{1-\epsilon}{2} \right )^2 + \gamma^2 \right]}
\label{equ11}
\end{eqnarray}

We can now extract information about the nature of eigenfunctions of the
original ladder network in different parametric and energy space by
considering the two Eqs.~(\ref{equ9}) and (\ref{equ10}) simultaneously.
\begin{figure}[ht]
{\centering\resizebox*{7cm}{3cm}{\includegraphics{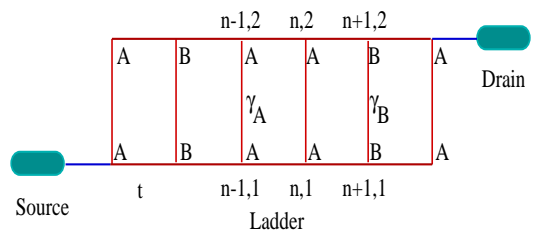}} \par}
\caption{(Color online). Schematic view of a ladder attached to two 
electrodes, where each arm of the ladder is a Fibonacci chain.}
\label{bridge1}
\end{figure}
First of all, for any fixed value of $\epsilon$, the set of equations
($9$) and ($10$) together  defines a ladder network consisting of random
on-site potentials occupying the arms $1$ and $2$. As both these equations
represent randomly disordered one dimensional chains, one expects Anderson
localization of all the electronic states provided $\lambda_1$ and
$\lambda_2$ are non-zero. The states of the ladder network will be
exponentially localized. However, there is a point of interest.

It is well known that, for a randomly disordered chain of length $L$
with $L >>1$, but not very large, one encounters a distribution of
localization lengths. The distribution is characterized by `local'
Lyapunov exponents which can be different for different eigenstates
of the disordered sample \cite{tosatti}. Only in the thermodynamic limit
one single exponent dominates the  distribution and one can talk of a
`unique' characteristic localization length. In the case of a ladder of
large but finite length Eqs.~(\ref{equ9}) and (\ref{equ10}) represent
chains having two different widths of disorder and hence, {\it two different
distributions} of localization lengths. One can however, simulate the
thermodynamic limit by averaging over various disorder configurations.
As a result of such averaging the `distribution' of localization lengths
will be dominated by one Lyapunov exponent, and we can talk of a
characteristic localization length of the disordered sample. Assuming this
is done, the Eqs.~(\ref{equ9}) and (\ref{equ10}) will represent chains with
two different (characteristic) lengths of localization.

Let $\xi_1$ and $\xi_2$ be these characteristic localization lengths
corresponding to Eqs.~(\ref{equ9}) and (\ref{equ10}) respectively, and let's
set, without any loss of generality, $\xi_1 < L < \xi_2$. As the Fermi energy
is swept through the eigenvalue spectrum corresponding to Eq.~(\ref{equ9}),
the ladder doesn't conduct, as $L > \xi_1$. On the other hand, as the Fermi
energy coincides with any of the eigenvalues corresponding to the spectrum
provided by Eq.~(\ref{equ10}), the ladder shows a finite conductance. Thus,
the finite ladder exhibits a transition from a non-conducting to a
conducting phase and thus shows a {\em switch-like} behavior. Of course, 
the value of the conductance in the second case may not always be high.
\vskip 0.05in
\noindent
{\em Case II:} $\epsilon_{n,1}$ are random, $\epsilon_{n,2}=\epsilon
\epsilon_{n,1}$ and $\gamma_n=\gamma \epsilon_{n,1}$, such that
$\lambda_2 = 0$.
\vskip 0.05in
This implies, $\gamma_n = \sqrt{(\epsilon_{n,1} \epsilon_{n,2})}$.
For this special choice of the parameters, Eqs.~(\ref{equ9}) and
(\ref{equ10}) read,
\begin{equation}
[E-(\epsilon_{n,1}+\epsilon_{n,2})] \phi_{n,1} = t (\phi_{n+1,1} +
\phi_{n-1,1} )
\label{equ12}
\end{equation}
\begin{equation}
E \phi_{n,2} = t ( \phi_{n+1,2} + \phi_{n-1,2} )
\label{equ13}
\end{equation}
This is an interesting case. Eq.~(\ref{equ12}) represents the eigenvalue
equation for a disordered chain with $\epsilon_{n,1}$ being uncorrelated
random potentials. Therefore, the electronic states represented by
$\phi_{n,1}$ are exponentially localized. On the other hand,
Eq.~(\ref{equ13}) represents a perfectly ordered chain with on-site
potential being equal to zero at each site. All the eigenstates represented
by $\phi_{n,2}$ are  extended and the system represented by the set of
Eqs.~(\ref{equ13}) has an absolutely continuous energy spectrum ranging
from $E=-2t$ to $E=2t$ in the thermodynamic limit. This implies that, in
the actual system, all the states beyond $|E|=2t$ will be localized
exponentially (courtesy, Eq.~(\ref{equ12})) and we get mobility edges at
$E=\pm 2t$. This presents an example where the existence of mobility edges
can be proven {\it analytically} in a low dimensional disordered system,
such as a two-chain ladder discussed here.
\vskip 0.05in
\noindent
{\em Case III:} A Fibonacci ladder
\vskip 0.05in
Let us now discuss a quasiperiodic version of the ladder, viz, a Fibonacci
ladder. Each arm of the ladder is a quasiperiodic Fibonacci 
chain~\cite{kohmoto}. A binary Fibonacci chain is composed of two 
`letters' $A$ and $B$ and the consecutive Fibonacci generations are grown 
following the substitution rules, $A \rightarrow AB$ and $B \rightarrow A$ 
with $A$ as the seed \cite{kohmoto}. The on-site potentials now assume values
$\epsilon_{A,j}$ and $\epsilon_{B,j}$ for an $A$-type or a $B$-type vertex
in the ladder, $j$ being the arm-index. The variety of vertices makes the
inter-ladder hopping $\gamma_n$ take up values $\gamma_A$ or $\gamma_B$
depending on whether it connects the $A-A$ vertices or the $B-B$ vertices
in the ladder in the transverse direction (Fig.~\ref{bridge1}).

We first discuss a special case again, in the spirit of our earlier Case I.
We choose $\epsilon=0$ and $\gamma_n = \gamma\epsilon_{n,1}$. It implies,
$\epsilon_{n,2}=\epsilon \epsilon_{n,1}=0$, but the ladder still retains its
quasiperiodic Fibonacci character. Eq.~(\ref{equ9}) and Eq.~(\ref{equ10})
retain their forms, but now with $\lambda_1 = (1+\sqrt{1+4\gamma^2})/2$ and,
$\lambda_2 = (1-\sqrt{1+4\gamma^2})/2$. As $\epsilon_{n,1}$ is taken to be
distributed in a Fibonacci sequence along arm number one of the ladder, each
of the equations (\ref{equ9}) and (\ref{equ10}) represents equations for two
independent Fibonacci chains. The eigenstates for each of them are typically
{\it critical} \cite{kohmoto}, exhibiting power law localization with a
multifractal distribution of the exponents. Thus, the spectrum of the
Fibonacci ladder will be composed only of such critical states and no
question of localization-delocalization transition arises.

Now, as a second case, we select $\epsilon_{A,2}=\epsilon_{A,1}$ and,
$\epsilon_{B,2}=\epsilon_{B,1}$, $\gamma_A=\sqrt{(\epsilon_{A,1}
\epsilon_{A,2})}=\epsilon_{A,1}$ and $\gamma_B=\sqrt{(\epsilon_{B,1}
\epsilon_{B,2})}=\epsilon_{B,1}$, which automatically makes $\epsilon=1$
and $\gamma=1$. The same set of Eqs.~(\ref{equ12}) and
(\ref{equ13}) are obtained. Now, $\epsilon_{n,1}+\epsilon_{n,2}$ is either
$2\epsilon_A$ or $2\epsilon_B$. That is, Eq.~(\ref{equ12}) represents a one
dimensional Fibonacci chain for all the single particle states are critical
\begin{figure}[ht]
{\centering\resizebox*{8cm}{8cm}{\includegraphics{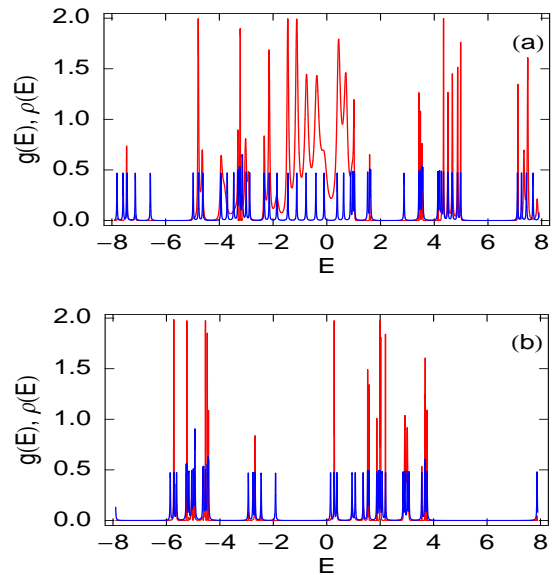}} \par}
\caption{(Color online). $g$-$E$ (red color) and $\rho$-$E$ (blue color) 
curves for a ladder
of total number of rungs $34$. (a) one chain (chain number $2$) is ordered
with on-site potential set equal to unity and the other chain (chain number
$1$) is subjected to Fibonacci modulation in site energies with
$\epsilon_{A,1}=-4$ and $\epsilon_{B,1}=4$ (b) both the chains are subjected
to Fibonacci modulation in site energies with $\epsilon_{A,1}=\epsilon_{A,2}
=-4$ and $\epsilon_{B,1}=\epsilon_{B,2}=4$. Other parameters are, $t_l=3$,
$\gamma=3$, and the on-site potential and the hopping integral in the
electrodes are set as $\epsilon_0=0$ and $t_0=4$ respectively. We have
chosen $c=e=h=1$.}
\label{fibo}
\end{figure}
\cite{kohmoto} and the spectrum in the thermodynamic limit, is a Cantor set,
with a gap in the neighborhood of every energy. The central part of the
spectrum of course, remains extended by virtue of Eq.~(\ref{equ13}),
refers to a perfectly ordered chain of atoms. Thus we again come across
mobility edges beyond $|E| = 2t$, but now its a transition from extended
to critical (power-law localized) states. The conductance accordingly
drops from (relatively) high to low values as one crosses such mobility
edges.

For a Fibonacci chain without the above restrictive values of the parameters,
we have to resort to numerical methods. Without any specified correlation
between the on-site potentials in the arms or in the values of the inter-arm
hopping $\gamma$, the decoupling of the ladder network into two independent
one dimensional chains is not possible (this is of course, true even with
the disordered ladder). However, the analysis made so far does not rule out
the possibility of a metal-insulator transition even in a general case.
As an example, we have performed a numerical calculation of the density of
states $\rho$ and conductance $g$ of a finite Fibonacci ladder. Results for
two separate cases are shown in Fig.~\ref{fibo}. In the first case, one arm
is an ordered chain while the other arm has a quasiperiodic Fibonacci
distribution of the on-site potentials. In the second case, both the arms
have a Fibonacci character.

For the numerical calculation we have adopted a Green's function formalism.
A finite ladder is attached to two semi-infinite one-dimensional perfect
electrodes, viz, source and drain, described by the standard tight-binding
Hamiltonian and parametrized by constant on-site potential $\epsilon_0$
and nearest neighbor hopping integral $t_0$ (already illustrated in
Fig.~\ref{bridge}). For low bias voltage and temperature, the conductance
$g$ of the ladder is determined by the single channel Landauer conductance
formula~\cite{datta} $g=(2e^2/h)T$. The transmission probability $T$ is
given by~\cite{datta} $T=Tr\left[\Gamma_S G_L^r \Gamma_D G_L^a\right]$.
$\Gamma_S$ and $\Gamma_D$ correspond to the imaginary parts of the
self-energies due to coupling of the ladder with the two electrodes and
$G_L$ represents the Green's function of the ladder~\cite{new,muj1,muj2}. 

In Fig.~\ref{fibo} we have superposed the picture of the density of states
on the conductance profile to show clearly that we have eigenstates existing
in energy regimes for which the conductance is very low. This illustrates
the transition from the conducting (high $g$) to non-conducting phase.

In conclusion, the results presented in this communication are worked out
for zero temperature. However, they should remain valid even in a certain
range of finite temperatures ($\sim 300$ K). This is because the broadening
of the energy levels of the ladder due to the electrode-ladder coupling is,
in general, much larger than that of the thermal broadening~\cite{datta}.
The inter ladder hopping $\gamma$ shifts the spectra
corresponding to Eqs.~(\ref{equ12}) and (\ref{equ13}) relative to each other.
As a result, in principle, one can tune the positions of the mobility
edges. This aspect may be inspiring in designing low-dimensional switching
devices, or even a DNA device.

\end{document}